\documentclass[aps,showpacs,preprintnumbers,amsmath,amssymb]{revtex4}
\usepackage{dcolumn}
\usepackage[dvips]{epsfig}
\usepackage{graphicx}
\usepackage{amssymb}
\usepackage{color}
\usepackage{enumerate}

\begin{document}
\baselineskip=0.8 cm
\title{\bf Testing gravity of a regular and slowly rotating phantom black hole by quasi-periodic oscillations}

\author{Songbai Chen$^{1}$\footnote{Corresponding author: csb3752@hunnu.edu.cn}, Mei Wang, Jiliang
Jing$^{1}$ \footnote{jljing@hunnu.edu.cn}}
%\email{csb3752@163.com}

\affiliation{Institute of Physics and Department of Physics, Hunan
Normal University,  Changsha, Hunan 410081, People's Republic of
China \\ Key Laboratory of Low Dimensional Quantum Structures \\
and Quantum Control of Ministry of Education, Hunan Normal
University, Changsha, Hunan 410081, People's Republic of China\\
Synergetic Innovation Center for Quantum Effects and Applications,
Hunan Normal University, Changsha, Hunan 410081, People's Republic
of China}

\begin{abstract}
\baselineskip=0.6 cm
\begin{center}
{\bf Abstract}
\end{center}

We extend firstly the regular phantom black hole solution to a slowly rotating black hole case and find that the phantom field depresses the angular velocity of the event horizon and suppresses the super-radiation of black hole.
We also probe the dependence of quasi-periodic oscillations frequencies in relativistic precession model on the phantom parameter. With the observation data of GRO J1655-40, we make a constraint on the parameters of the regular and slowly rotating phantom black hole. Our results show that although the best-fit value of the phantom parameter $b$ is small, the allowed value of $b$  in the $1\sigma$ region is $b<0.619$, which means that the phantom theoretical model can not be excluded by the constraint from quasi-periodic oscillations with the observation data of GRO J1655-40.

\end{abstract}

\pacs{ 04.70.Dy, 95.30.Sf, 97.60.Lf } \maketitle
\newpage
\section{Introduction}

Phantom dark energy is a special kind of theoretical models with the
negative kinetic energy\cite{Caldwell}, which has been investigated extensively in cosmology because it can provide a mechanism to interpret the accelerating expansion of the current Universe \cite{ph1,ph2,ph3,ph4,ph5,ph6,ph61,ph62,ph63,ph64,ph65}. Due to its negative kinetic energy, the phantom dark energy owns the super negative equation of state $w<-1$, which leads to that the null energy condition is violated for phantom field. The Universe dominated by phantom dark energy will blow up incessantly and arrive at a future singularity named big rip at where anything in the universe including the large galaxies will be torn up. Although the phantom dark energy owns such exotic properties, it is still supported  by recent precise observational data \cite{ph7,ph71}. However, all of these observational examinations and constraints on phantom dark energy are taken in the scope of Cosmology.

In black hole physics, the behaviors and properties of phantom field have been studied in last few years. E. Babichev \cite{EBab} found that after absorbing
phantom field, the mass of black hole decreases, which implies that the cosmic censorship conjecture could be challenged severely since the charge of a Reissner-Nordstr\"{o}m-like black hole could be larger than its mass as it accretes the phantom energy. We studied dynamical evolution of the phantom scalar perturbation in the Schwarzschild black hole spacetime and found that the phantom scalar perturbation grows with an exponential
rate in the late-time evolution, which differs from the decay of the usual scalar perturbations in the background of a black hole \cite{Sb,Sb1,bw}. Moreover, some black hole solutions describing gravity coupled to phantom scalar fields or phantom Maxwell fields have been found  and the corresponding geometric structure and thermodynamic properties are also studied in \cite{pbh1,pbh11,pbh12,pbh2,pbh3,pbh4,pbh5,pbh6,pbh8,pbh9}. The strong gravitational lensing of such kind of black hole with phantom hair has been investigated in \cite{pbh7,GL1,GL2,GL3,GL4}. These investigations discloses some effects of phantom field on the black hole physics. However, the observation examinations and constraints on phantom dark energy from black hole physics are still lacking.

Quasi-periodic oscillations is one of very promising tools to test theories of gravity in the strong field regime and to constrain black hole parameters. The investigations show that quasi-periodic oscillations are a common feature in the X-ray power density
spectrum of black hole binaries \cite{qpo1,qpo2}. According to the types and the properties of quasi-periodic oscillations, their low frequencies are distributed in the range $0.1\sim 30$ Hz and high-frequencies are in the range $100\sim500$ Hz.  The relativistic precession model is proposed as an exact mechanism to explain twin high-frequency quasi-periodic oscillations as well as a low-frequency mode in low-mass X-ray binaries \cite{RPM1,RPM2,RPM20,RPM3,RPM31,RPM32,RMP4,RMP5}. In this precession model, the twin higher frequencies are identified respectively with the azimuthal frequency $\nu_{\phi}$ and the periastron precession frequency $\nu_{\text{per}}$ of a test-particle moving in quasi-circular orbits at the innermost disk region in the background spacetime. The low-frequency mode in quasi-periodic oscillations is regarded as the nodal precession frequency $\nu_{\text{nod}}$, which is emitted at the same radius where the twin higher frequencies signals are generated. The possibility of applying quasi-periodic oscillations to constrain the black hole parameters in various theories of gravity has been already investigated in \cite{TB1,TB11,TB2,TB3,TB5,TB6,TB7,TB8}.

The main purpose of this paper is to make a constraint on the parameters of a regular and slowly rotating phantom black hole by using of quasi-periodic oscillations in relativistic precession model and the observation data of GRO J1655-40. Meanwhile, we want to see whether phantom theoretical model can pass an observation test from black hole physics.

The paper is organized as follows: in the following section we will extend the regular phantom black hole solution \cite{pbh3} to the slowly rotating black hole case and then study the effect of the phantom parameter on the angular velocity at the event horizon. In Sec.III, we make a constraint on the parameters of a regular and slowly rotating phantom black hole by using of quasi-periodic oscillations and the observation data of GRO J1655-40. Finally, we end the paper with a summary.

\section{A regular and slowly rotating phantom black hole}

In this section, we will extend the regular phantom black hole solution to the slowly rotating black hole case. Let us first introduce briefly the regular and static phantom black hole obtained in literature \cite{pbh3}. The action for phantom scalar field $\Phi$ in the curve spacetime is
\begin{eqnarray}
S=\int \sqrt{-g}d^4x[R-\frac{1}{2}g^{\mu\nu}\partial_{\mu}\Phi\partial_{\nu}\Phi+V(\Phi)].\label{act1}
\end{eqnarray}
After solving the equation of motion of phantom scalar field and Einstein field equation, Bronnikov \textit{et al} \cite{pbh3} obtained a regular and static phantom black hole described by
\begin{eqnarray}
ds^2=-f(r)dt^2+\frac{1}{f(r)}dr^2+(r^2+b^2)(d\theta^2+r^2\sin^2{\theta}
d\phi^2), \label{metric0}
\end{eqnarray}
with
\begin{eqnarray}
f(r)&=&1-\frac{3M}{b}\bigg[(\frac{\pi}{2}-\arctan\frac{r}{b})
(1+\frac{r^2}{b^2})-\frac{r}{b}\bigg],\label{f0}\\
\Phi&\equiv&\sqrt{2}\psi=\sqrt{2} \arctan \frac{r}{b},\\
V&=&\frac{3M}{b^3}\bigg[(\frac{\pi}{2}-\psi)(3-2\cos^2\psi)-3\sin\psi\cos\psi\bigg].
\label{ff1}
\end{eqnarray}
Here $M$ is the mass of black hole and $b$ is a positive constant related to the charge of phantom scalar field. When $b$ tends to zero, the phantom scalar field $\Phi$ becomes a constant and the corresponding potential $V$ approaches to zero, which means the action reduces to the usual action without material field and then the black hole solution (\ref{metric0}) recovers the known Schwarzschild black hole one. For the phantom black hole spacetime with non-zero value of $b$, it is regular and has no singularity because the curvature scalar
\begin{eqnarray}
R_{\mu\nu\rho\tau}R^{\mu\nu\rho\tau}=&&\frac{1}{b^6(r^2+b^2)^4}\bigg\{
108M^2(r^2+b^2)^2 (2r^4+2b^2r^2+b^4)\bigg(\frac{\pi}{2}-\arctan\frac{r}{b}\bigg)^2\nonumber\\&&
-24bM(r^2+b^2)\bigg(\frac{\pi}{2}-\arctan\frac{r}{b}\bigg)\bigg[b^4(r^2+2b^2) +3Mr(6r^4+10b^2r^2+5b^4)\bigg]\nonumber\\&&
+36b^2M^2r^2(6r^4+14b^2r^2+11b^4)+12b^6(2Mr^3+8b^2Mr+b^4)\bigg\},
\end{eqnarray}
is neither divergent nor zero at anywhere. Near the region $b\sim 0$, one can find
\begin{eqnarray}
R_{\mu\nu\rho\tau}R^{\mu\nu\rho\tau}=\frac{48M^2}{r^6}+\frac{32(r-6M)b^2}{r^8}+O(b^4),
\end{eqnarray}
which means that as the phantom charge $b$ disappears $R_{\mu\nu\rho\tau}R^{\mu\nu\rho\tau}$ of the regular phantom black hole spacetime can reduce to that of Schwarzschild black hole one.

We are now in position to obtain a phantom rotating black hole solution. For the electro-vacuum case, one can obtain a rotating black hole solution from a static black hole one by the Newman-Janis algorithm \cite{NJ1}. For example,
one can obtain Kerr black hole solution from Schwarzschild one by this technique. However, for the cases with scalar field, one must improve the usual Newman-Janis algorithm so that the rotating counterpart obeys to Einstein equation with scalar field. In general, it is difficult how to improve the Newman-Janis technique for the non-electro-vacuum cases. Therefore, we here focus on obtaining a regular slowly rotating phantom
black hole solution originated from the static and spherical symmetric solution (\ref{metric0}) by solving Einstein equation of the gravity system (\ref{act1}).

The metric of a regular slowly rotating phantom
black hole can be assumed as
\begin{eqnarray}
ds^2&=&-U(r)dt^2+\frac{1}{U(r)}dr^2-2F(r,\theta)adtd\phi+(r^2+b^2)(d\theta^2+\sin^2{\theta}
d\phi^2), \label{metric3}
\end{eqnarray}
where $a$ is the rotation parameter associated with its angular momentum. In the case of a slowly rotating spacetime (\ref{metric3}), we can suppose both of the phantom scalar field $\Phi$ and the metric function $U(r)$ depend only on the radial coordinate $r$. And then the equation of the phantom scalar field in spacetime (\ref{metric3}) can be expressed as
\begin{eqnarray}
\frac{1}{r^2+b^2}\frac{d}{dr}\bigg[(r^2+b^2)U(r)\Phi'\bigg]
-\frac{dV(\Phi)}{d\Phi}=0,
\end{eqnarray}
which is similar to that in the static and spherical symmetric
spacetime (\ref{metric0}) because the phantom scalar field $\Phi$ is supposed to only a function of $r$ and all of the higher order terms $\mathcal{O}(a^2)$ are neglected.

Inserting the metric (\ref{metric3}) into the Einstein's field equation, we find that the non-vanishing components of field equation can be expanded to first order in the angular momentum parameter $a$ as
\begin{eqnarray}
tt: &&\frac{U(r)}{(r^2+b^2)^2}\bigg[U'(r)r(r^2+b^2)+U(r)(r^2+2b^2)-(r^2+b^2)\bigg]
=\frac{1}{2}[U(r)\Phi'^2-2V]+\mathcal{O}(a^2),\label{gt1}\\
rr:
&&\frac{1}{U(r)(r^2+b^2)^2}\bigg[U'(r)r(r^2+b^2)+U(r)r^2-(r^2+b^2)\bigg]
=-\frac{1}{2U(r)}[U(r)\Phi'^2+2V]+\mathcal{O}(a^2)\label{gt2},\\
\theta\theta:&&\frac{1}{2(r^2+b^2)}\bigg[U''(r)(r^2+b^2)^2+2U'(r)r(r^2+b^2)
+2U(r)b^2\bigg]=\frac{1}{2}(r^2+b^2)[U(r)\Phi'^2-2V]+\mathcal{O}(a^2)\label{gt3},\\
\phi\phi:&&\frac{\sin^2\theta}{2(r^2+b^2)}\bigg[U''(r)(r^2+b^2)^2+2U'(r)r(r^2+b^2)
+2U(r)b^2\bigg]=\frac{\sin^2\theta}{2}(r^2+b^2)[U(r)\Phi'^2-2V]+\mathcal{O}(a^2)\nonumber,\\
t\phi:&&\frac{1}{2(r^2+b^2)^2}\bigg\{(r^2+b^2)^2U(r)\frac{\partial^2F(r,\theta)}{\partial
r^2}-F(r,\theta)\bigg[(r^2+b^2)^2U''(r)+2rU'(r)(r^2+b^2)+2U(r)(r^2+2b^2)\bigg]\nonumber\\&&
+(r^2+b^2)\bigg[2F(r,\theta)+
\frac{\partial^2F(r,\theta)}{\partial \theta^2}-\frac{\partial
F(r,\theta)}{\partial \theta}\cot\theta\bigg]\bigg\}+
\frac{F(r,\theta)}{2}[U(r)\Phi'^2-2V]=0+\mathcal{O}(a^2)\label{gt5}.
\end{eqnarray}
Solving the Einstein equations (\ref{gt1})-(\ref{gt3}), one can obtain the metric function
\begin{eqnarray}
U(r)=f(r)=1-\frac{3M}{b}\bigg[(\frac{\pi}{2}-\arctan\frac{r}{b})
(1+\frac{r^2}{b^2})-\frac{r}{b}\bigg].\label{fba22}
\end{eqnarray}
Separating $F(r,\theta)=h(r)\Theta(\theta)$, we find that the angular part $\Theta(\theta)$ satisfies
\begin{eqnarray}
\frac{d^2\Theta(\theta)}{d\theta^2}-\frac{d\Theta(\theta)}{d\theta}\cot\theta
+\lambda\Theta(\theta)=0,\label{angu}
\end{eqnarray}
and the radial part $h(r)$ in Eq.(\ref{gt5}) obeys to
\begin{eqnarray}
(r^2+b^2)U(r)\frac{d^2h(r)}{dr^2}-2h(r)\bigg[U(r)-1+\frac{\lambda}{2}\bigg]=0,\label{radial}
\end{eqnarray}
where $\lambda$ is a separation constant.
In order that the coefficient $g_{t\phi}$ can be reduced to that in the usual slowly rotating black hole without phantom field, here we set $\lambda=2$ and then find that $\Theta(\theta)=\sin^2\theta$ in this
case. This yields that the radial part of equation (\ref{radial}) can be simplified as
\begin{eqnarray}
(r^2+b^2)\frac{d^2h(r)}{dr^2}-2h(r)=0,
\end{eqnarray}
which leads to
\begin{eqnarray}
h(r)= \frac{3M}{b}\bigg[(\frac{\pi}{2}-\arctan\frac{r}{b})
(1+\frac{r^2}{b^2})-\frac{r}{b}\bigg].
\label{hr}
\end{eqnarray}
Like in a usual slowly rotating black hole, the event horizon of black hole (\ref{metric3}) is given by $f(r)=0$, which is
the same as that in the static and spherical symmetric case
(\ref{metric0}) since we here expand the metric
only to first order in the angular momentum parameter $a$. When the parameter $b\rightarrow 0$, one can find that $h(r)\rightarrow \frac{2M}{r}$, which recovers that of a usual slowly
rotating black hole without phantom scalar field. The
angular velocity of the horizon $\Omega_H$ is an important quantity for a rotating black hole, which affects the region at where the super-radiance occurs in the black hole background. In
the spacetime of a regular and slowly rotating phantom black hole (\ref{metric3}), the angular velocity $\Omega_H$ has a form
\begin{eqnarray}
\Omega_H=-\frac{g_{t\phi}}{g_{\phi\phi}}\bigg|_{r=r_H}=\frac{a}{r^2_H+b^2},
\end{eqnarray}
\begin{figure}[ht]
\begin{center}
\includegraphics[width=7cm]{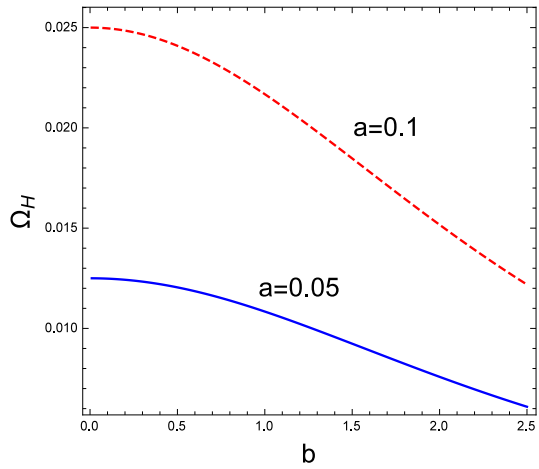}
\caption{The change of the angular velocity $\Omega_H$ with the
parameter $b$ in a regular and slowly rotating phantom black hole spacetime.}
\end{center}
\end{figure}
which depends on the phantom charge $b$. We
plot the change of the angular velocity $\Omega_H$ with the
parameter $b$ in Fig.(1), which tells us that $\Omega_H$
decreases monotonically with the phantom charge $b$ and then the super-radiation is suppressed by phantom field in this case.

\section{Constraint on parameters of a regular and slowly rotating phantom black hole by quasi-periodic oscillations}

In this section, we will make a constraint on parameters of above regular and slowly rotating phantom black hole by quasi-periodic oscillations. In the general stationary and axially symmetric spacetime
\begin{eqnarray}
ds^2&=&g_{tt}dt^2+g_{rr}dr^2+2g_{t\phi}dtd\phi+g_{\theta\theta}d\theta^2
+g_{\phi\phi}
d\phi^2, \label{metric3n}
\end{eqnarray}
one can find that for the geodesic motion of particle there exist
the conserved specific energy at infinity $E$ and the conserved $z$-component
of the specific angular momentum at infinity $L_z$ since
the metric coefficients are independent of the coordinates $ t$ and $\phi$.
And then the timelike geodesics can be expressed as
\begin{eqnarray}
&&\dot{t}=\frac{Eg_{\phi\phi}+L_zg_{t\phi}}{g^2_{t\phi}-g_{tt}g_{\phi\phi}},\label{u1}\\
&&\dot{\phi}=-\frac{Eg_{t\phi}+L_zg_{tt}}{g^2_{t\phi}-g_{tt}g_{\phi\phi}},\label{u2}\\
&&g_{rr}\dot{r}^2+g_{\theta\theta}\dot{\theta}^2=V_{eff}(r,\theta; E,L_z),\label{u3}
\end{eqnarray}
with the effective potential
\begin{eqnarray}
V_{eff}(r)=\frac{E^2g_{\phi\phi}+2EL_zg_{t\phi}+L^2_zg_{tt}
}{g^2_{t\phi}-g_{tt}g_{\phi\phi}}-1,
\end{eqnarray}
where the overhead dot represents a derivative with respect to the
affine parameter. For a
circular orbit in the equatorial plane $\theta=\pi/2$, we have
\begin{eqnarray}
V_{eff}(r)=0, \;\;\;\;\;\;\;\frac{dV_{eff}(r)}{dr}=0.
\end{eqnarray}
Solving above equations, one can obtain
\begin{eqnarray}
&&E=-\frac{g_{tt}+g_{t\phi}\Omega_{\phi}}{\sqrt{-g_{tt}-2g_{t\phi}\Omega_{\phi}
-g_{\phi\phi}\Omega^2_{\phi}}},\nonumber\\
&&L_z=\frac{g_{t\phi}+g_{\phi\phi}\Omega_{\phi}}{\sqrt{g_{tt}
+2g_{t\phi}\Omega_{\phi}-g_{\phi\phi}\Omega^2_{\phi}}},\nonumber\\
&&\Omega_{\phi}=\frac{d\phi}{dt}=\frac{-g_{t\phi,r}+\sqrt{(g_{t\phi,r})^2
+g_{tt,r}g_{\phi\phi,r}}}{g_{\phi\phi,r}},\label{jsd}
\end{eqnarray}
where $\Omega_{\phi}=2\pi\nu_{\phi}$ is the angular velocity of particle moving in the circular orbits and $\nu_{\phi}$ is its corresponding azimuthal frequency. Considering a small perturbation of a circular,
equatorial orbit, i.e.,
\begin{eqnarray}
r(t)=r_0+\delta r(t), \;\;\;\;\;\;\;\;\;\;\theta(t)=\frac{\pi}{2}+\delta \theta(t),
\end{eqnarray}
one can find that the perturbations $\delta r(t)$ and $\delta \theta(t)$ are governed by the following differential equations
\begin{eqnarray}
\frac{d^2\delta r(t) }{dt^2}+\Omega^2_{r}\delta r(t)=0, \;\;\;\;\;\;\;\;\;\;\frac{d^2\delta \theta(t) }{dt^2}+\Omega^2_{\theta}\;\delta \theta(t)=0,
\end{eqnarray}
with
\begin{eqnarray}
\Omega^2_{r}=-\frac{1}{2g_{rr}\dot{t}^2}\frac{\partial^2 V_{eff}}{\partial r^2}, \;\;\;\;\;\;\;\;\;\;\Omega^2_{\theta}=-\frac{1}{2g_{\theta\theta}\dot{t}^2}\frac{\partial^2 V_{eff}}{\partial \theta^2},
\end{eqnarray}
and
 \begin{eqnarray}
\dot{t}=\frac{1}{\sqrt{-g_{tt}-2g_{t\phi}\Omega_{\phi}
-g_{\phi\phi}\Omega^2_{\phi}}}.
\end{eqnarray}
The radial epicyclic frequency $\nu_r$ and the vertical
epicyclic frequency $\nu_{\theta}$ can be written as $\nu_r=\Omega_r/2\pi$ and $\nu_{\theta}=\Omega_{\theta}/2\pi$, respectively.
Finally, the periastron and nodal precession frequencies can be expressed as
\begin{eqnarray}
\nu_{\text{per}}=\nu_{\phi}-\nu_{r},\;\;\;\;\;\;\;\;\;\;\;
\nu_{\text{nod}}=\nu_{\phi}-\nu_{\theta}.
\end{eqnarray}
\begin{figure}[ht]
\begin{center}
\includegraphics[width=5.4cm]{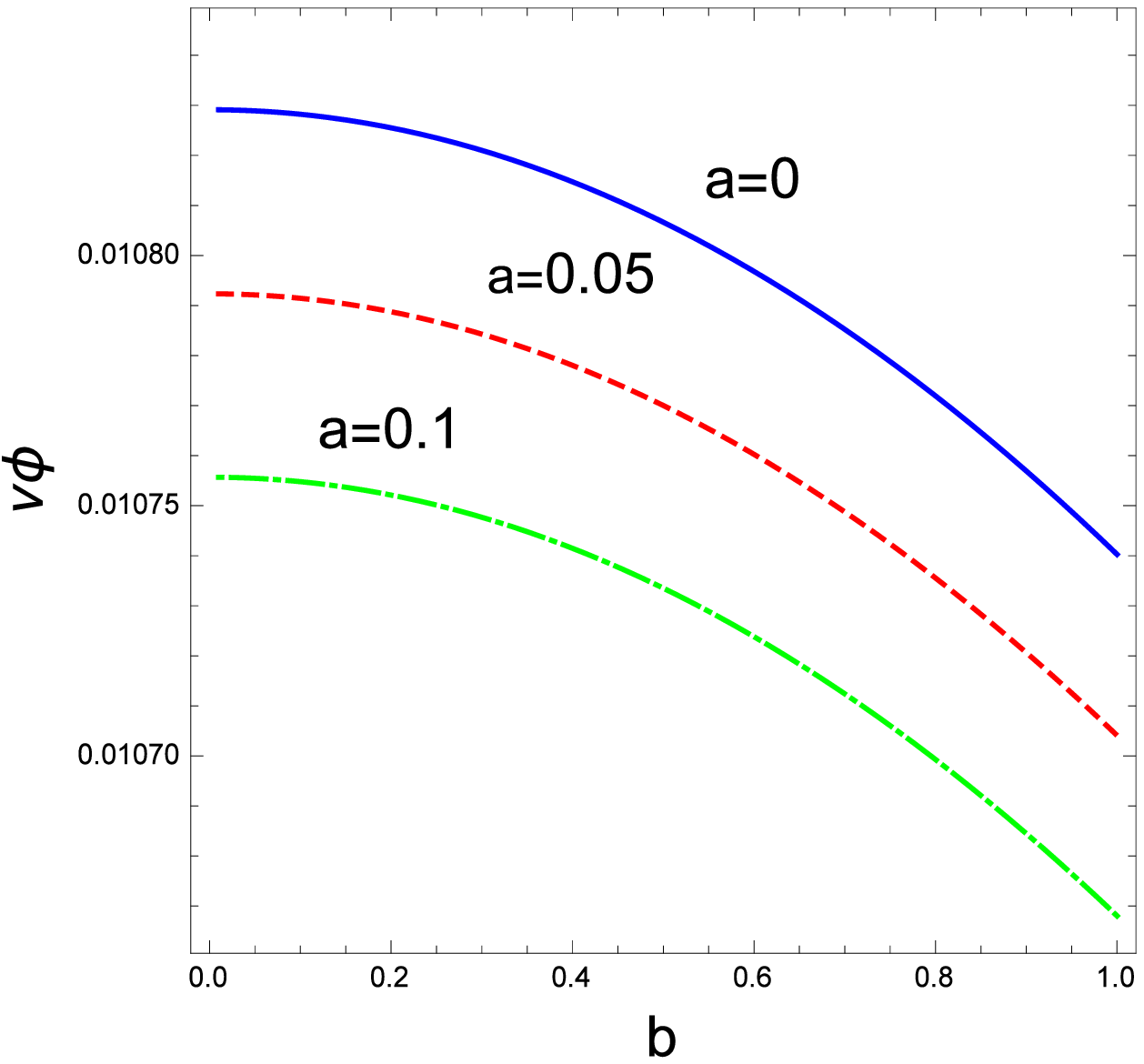}\;\includegraphics[width=5.4cm]{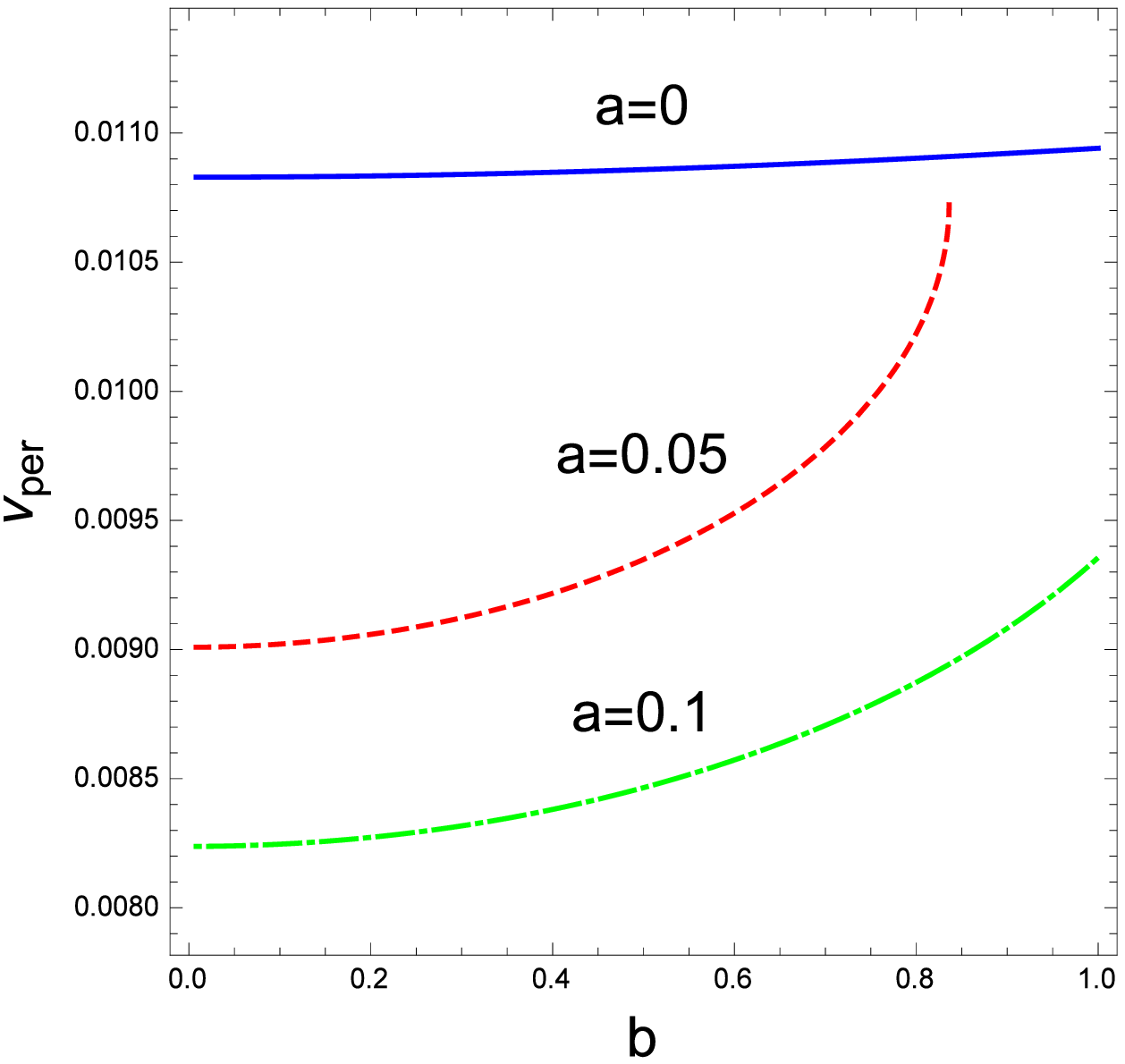}
\;\includegraphics[width=5.4cm]{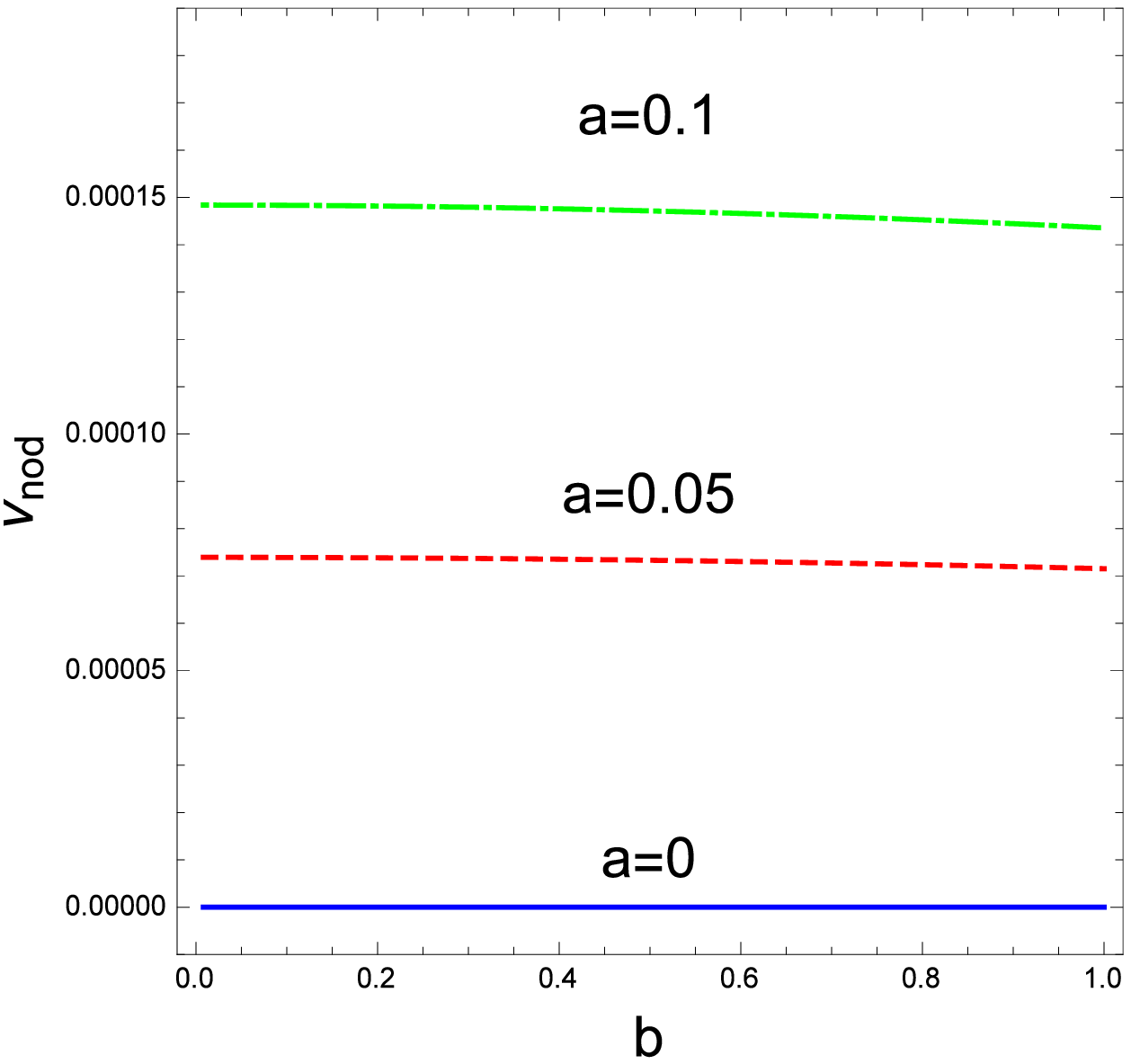}
\caption{The change of the frequencies $\nu_{\phi}$, $\nu_{\text{per}}$ and $\nu_{\text{nod}}$ with the parameter $b$ in a regular and slowly rotating phantom black hole spacetime. Here we set $M=1$ and $r=6$.}
\end{center}
\end{figure}
In Fig.(2), we plot the dependence of the frequencies $\nu_{\phi}$, $\nu_{\text{per}}$ and $\nu_{\text{nod}}$ on the parameter $b$ in a regular and slowly rotating phantom black hole spacetime. With the increase of the phantom charge $b$, the azimuthal frequency $\nu_{\phi}$ decreases and the periastron precession frequency $\nu_{\text{per}}$ increases. For the non-zero $a$ case, the nodal precession frequency $\nu_{\text{nod}}$ also decreases with the parameter $b$. However, for the non-rotation case, the nodal precession frequency $\nu_{\text{nod}}$ becomes zero and is independent of the phantom parameter $b$, which means that the vertical epicyclic frequency $\nu_{\theta}$ is identical to the azimuthal frequency $\nu_{\phi}$ for the spacetime (\ref{metric3}) with $a=0$.

If three quasi-periodic oscillations frequencies are observed simultaneously, one can associate them to orbital and precession frequencies in the relativistic precession model. For the usual Kerr black hole spacetime,
one can solve the corresponding three equations and determine the three variables ($r$, $M$, and $a$). However, for the regular and slowly rotating phantom black hole, there are extra parameter $b$ related to the phantom scalar field, which means that we can not get values of the four variables through solving three equations directly. Here, we must resort to  the $\chi^2$ analysis to best-fit the values of four unknown variables ($r$, $M$,  $a$ and $b$ ) in the phantom black hole spacetime (\ref{metric3}) with the three observed frequencies. From the current observations of GRO J1655-40, there are two set of data about these frequencies ($\nu_{\phi}, \nu_{\text{per}},\nu_{\text{nod}}$ )\cite{RPM1}:
\begin{eqnarray}
(441^{+2}_{-2},\;\;298^{+4}_{-4}, \;\;17.3^{+0.1}_{-0.1})\;\;\;\; \text{and}\;\;\;\;
(451^{+5}_{-5},\;\;-, \;\;18.3^{+0.1}_{-0.1}),
\end{eqnarray}
Moreover, there also is an independent dynamical measurement
of the mass of the black hole \cite{TB4}: $M_{\text{dyn}}=5.4\pm0.3M_{\odot}$. Therefore, there are five free parameters: mass $M$, rotation parameter $a$, the phantom scalar charge $b$,
the radius $r_1$ and $r_2$ correspond the observations with three frequencies and two frequencies, respectively.
With these data, the gravity of Kerr black hole was tested in the relativistic precession model \cite{TB1}. For the phantom black hole (\ref{metric3}), we obtain the minimum $\chi^2_{\text{min}}=0.3329$ and constrain the black hole parameters
\begin{eqnarray}
M=5.274^{+0.054}_{-0.053}M_{\odot},\;\;\;\;\;\;\;\;a^{*}\equiv a/M=0.258^{+0.005}_{-0.003},
\;\;\;\;\;\;\;\;b=0.044^{+0.575}_{-0.044},
\end{eqnarray}
at the $68.3\%$ confidence level.
\begin{figure}[ht]
\begin{center}
\includegraphics[width=7cm]{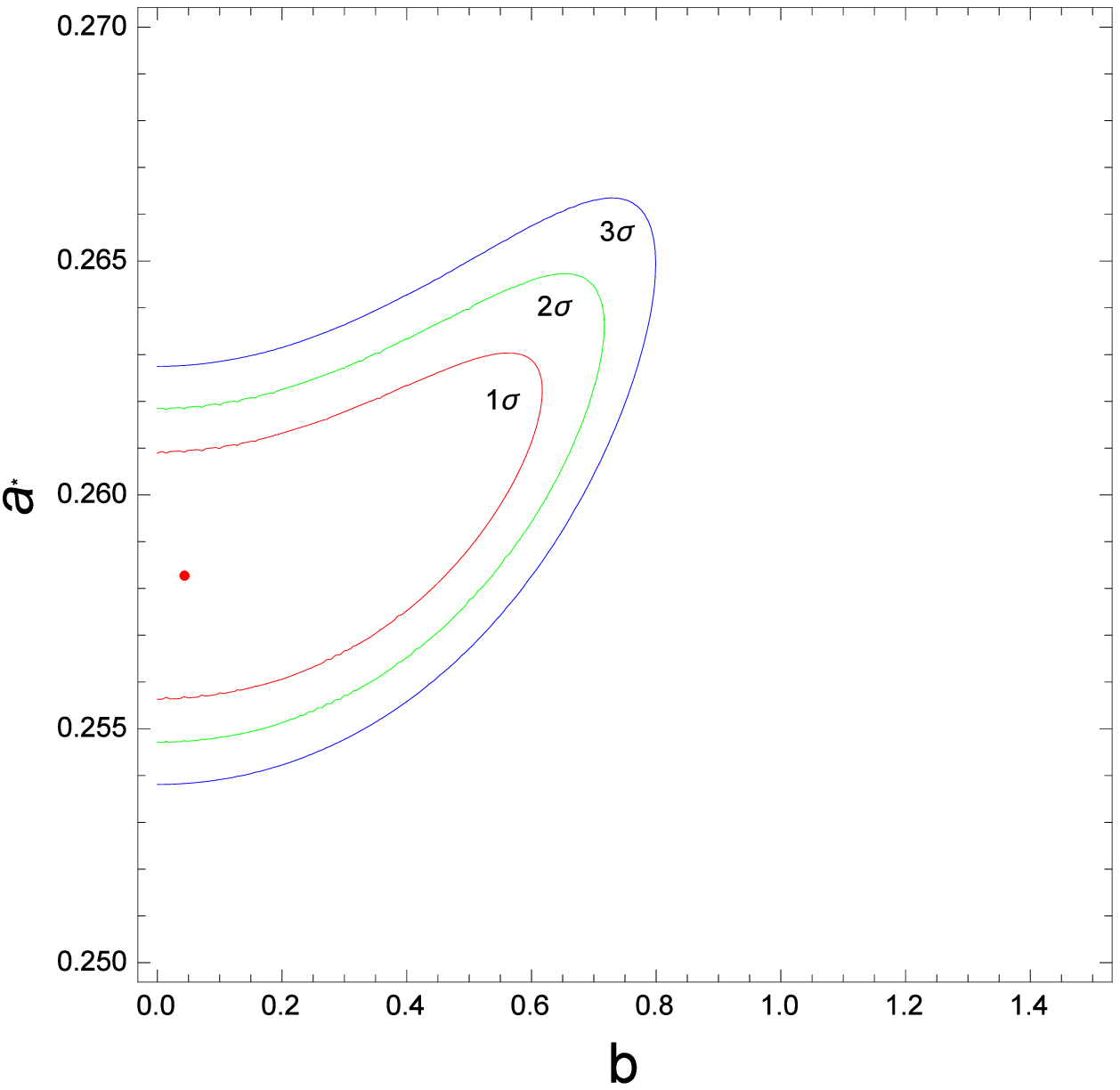}\;\;\;\;\includegraphics[width=7cm]{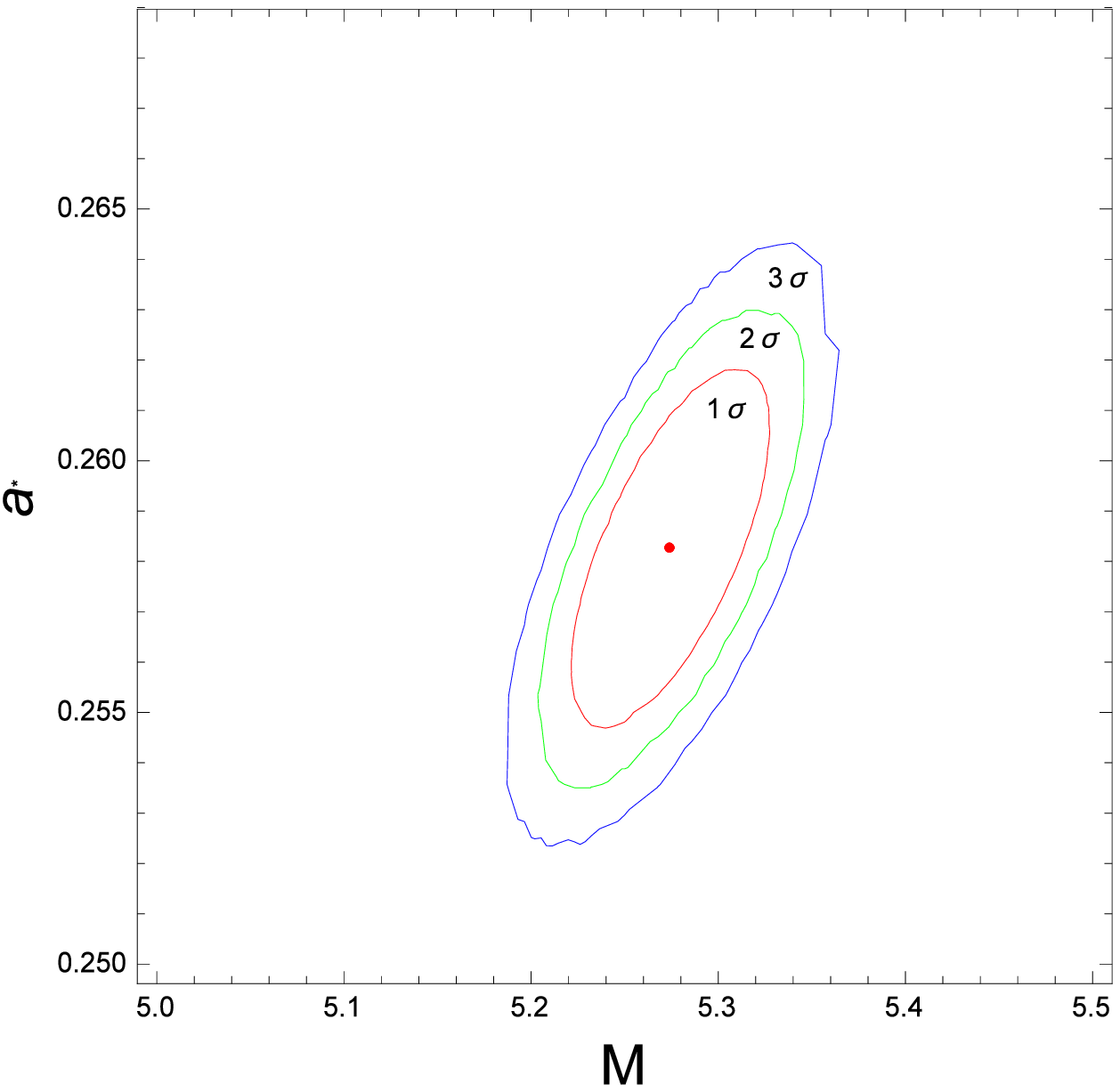}
\caption{Constraints on the parameters of a regular and slowly rotation phantom black hole with the black hole candidate in GRO J1655-40 from current observations of QPOs within the relativistic precession model. The red, green and blue lines represent the contour levels $1\sigma$, $2\sigma$ and $3\sigma$, respectively. The red dot in the panels correspond the best-fit values of parameters: $M=5.274$, $a^*=a/M=0.258$ and $b=0.044$.}
\end{center}
\end{figure}
 The best-fit values of the radius of circular orbital corresponding two sets of quasi-periodic oscillations are $r_1=5.717 M=1.129 \;r_{\text{ISCO}}$ and $r_2=5.614 M=1.1086 \;r_{\text{ISCO}}$, respectively. Here $r_{\text{ISCO}}$ is the innermost stable circular orbit in the regular and slowly rotating phantom black hole spacetime with the best-fit values ($M=5.274$, $a^*=a/M=0.258$ and $b=0.044$). It means that the circular orbit  of quasi-periodic oscillations are located at the strong gravitational-field region of the black hole because the black hole event horizon is at $1.9998M$. Moreover, the contour levels of $1\sigma$, $2\sigma$ and $3\sigma$ for the mass $M$, rotation parameter $a$ and the phantom scalar charge $b$ are shown in Fig.(3). From the left panel in Fig.(3), we find that the contour lines with levels of $1\sigma$, $2\sigma$ and $3\sigma$ are not closed curves since the parameter $b$ is non-negative for the regular phantom black holes (\ref{metric0}) and (\ref{metric3}). Our results show that although the best-fit value of the phantom parameter $b$ is small, the allowed region of $b$ at the $68.3\%$ confidence level is $b<0.619$, which means that the deviations
from Kerr metric is possible and the phantom theoretical model can not be excluded by the constraint from quasi-periodic oscillations with the observation data of GRO J1655-40.

\section{summary}

In this paper we extend firstly the regular phantom black hole solution to the slowly rotating case and find that the presence of phantom field depresses the angular velocity of the event horizon and suppresses the super-radiation of black hole. And then, we study the dependence of  quasi-periodic oscillations frequencies in relativistic precession model on the phantom parameter $b$.
With the increase of the phantom charge $b$, the azimuthal frequency $\nu_{\phi}$ decreases and the periastron precession frequency $\nu_{\text{per}}$ increases.  The nodal precession frequency $\nu_{\text{nod}}$ also decreases with the parameter $b$ for the non-zero $a$ case, but it becomes zero in the case $a=0$. With the observation data of GRO J1655-40, we constrain the parameters of the regular and slowly rotating phantom black hole. Our results show that although the best-fit value of the phantom parameter $b$ is small, the allowed region of $b$ at the $68.3\%$ confidence level is $b<0.619$, which means that it is possible for a black hole has a phantom scalar hair. In other words, the phantom theoretical model can not be excluded by the constraint from quasi-periodic oscillations with the observation data of GRO J1655-40.

\section{\bf Acknowledgments}

This work was partially supported by the National Natural
Science Foundation of China under Grant No.11275065,  No. 11475061,
the construct program of the National Key Discipline, and the Open
Project Program of State Key Laboratory of Theoretical Physics,
Institute of Theoretical Physics, Chinese Academy of Sciences, China
(No.Y5KF161CJ1).


\vspace*{0.2cm}
\begin{thebibliography}{99}

\baselineskip=0.6 cm \baselineskip=0.6 cm

\bibitem{Caldwell} R. R. Caldwell, Phys. Lett. B {\bf545}, 23 (2002), arXiv: astro-ph/9908168.

\bibitem{ph1} B. McInnes, J. High Energy Phys. {\bf08}, 029 (2002),  arXiv: hep-th/0112066.

\bibitem{ph2} S. Nojiri and S. D. Odintsov Phys. Lett. B, {\bf562} 147(2003), arXiv:hep-th/0303117.

\bibitem{ph3} L. P. Chimento and R. Lazkoz Phys. Rev. Lett. {\bf91} 211301 (2003),arXiv:gr-qc/0307111.

\bibitem{ph4} B. Boisseau, G. Esposito-Farese, D. Polarski and A. A. Starobinsky, Phys. Rev. Lett. {\bf85}, 2236 (2000), arXiv:gr-qc/0001066.

\bibitem{ph5} Gannouji R, Polarski D, Ranquet A and Starobinsky A A 2006
\textit{J. Cosmol. Astropart. Phys.} {\bf09} 016


\bibitem{ph6} R. R. Caldwell, M. Kamionkowski  and N. N. Weinberg,
Phys. Rev. Lett. {\bf91} 071301 (2003).
\bibitem{ph61} S. Nesseris and L. Perivolaropoulos, Phys. Rev. D {\bf70}, 123529 (2004), arXiv:astro-ph/0410309.
\bibitem{ph62} S. Nojiri and S. D. Odintsov, Phys. Lett. B {\bf571}, 1 (2003), arXiv:hep-th/0306212.
\bibitem{ph63} P. Singh, M. Sami and N. Dadhich, Phys. Rev. D {\bf68}, 023522 (2003).
\bibitem{ph64} J. G. Hao and  X. Z. Li, Phys. Rev. D {\bf70} 043529 (2004), arXiv: astro-ph/0309746.
\bibitem{ph65} E. N. Saridakis,  P. F. Gonzalez-Diaz and C. L. Siguenza, Class. Quant. Grav. {\bf26}, 165003 (2009).


\bibitem{ph7} A. Melchiorri, L. Mersini-Houghton, C. J. Odman and M. Trodden, Phys. Rev. D {\bf68}, 043509  (2003), arXiv:astro-ph/0211522.
\bibitem{ph71} M. A.  Ainou, Phys. Rev. D {\bf87}, 024012 (2013), arXiv: 1209.5232.


\bibitem{EBab} E. Babichev, V. Dokuchaev  and Y. Eroshenko,  Phys. Rev. Lett. {\bf93} 021102 (2004).


\bibitem{Sb} S. Chen, J. Jing and Q. Pan, Phys. Lett. B {\bf670}, 276  (2009).
 \bibitem{Sb1}S. Chen and J. Jing, J. High Energy Phys. {\bf03}, 081  (2009).
\bibitem{bw} X. He, B. Wang, S. Wu and C. Lin, Phys. Lett. B {\bf673}, 156 (2009).


\bibitem{pbh1} G. W. Gibbons and D. A. Rasheed, Nucl. Phys. B {\bf476} 515  (1996),(arXiv:hep-th/9604177).
\bibitem{pbh11} G. Cl\'{e}ment, J. C. Fabris and M. E. Rodrigues, Phys. Rev. D {\bf79} 064021 (2009), arXiv:0901.4543.
\bibitem{pbh12}M. Azreg-Ainou, G. Cl\'{e}ment, J. C. Fabris and M. E. Rodrigues, Phys. Rev. D {\bf83} 124001 (2011), arXiv:1102.4093.

\bibitem{pbh2} C. J. Gao and S. N. Zhang, arXiv:hep-th/0604114.

\bibitem{pbh3} K. A. Bronnikov and J. C. Fabris, Phys. Rev. Lett. {\bf96} 251101 (2006), arXiv:gr-qc/0511109.

\bibitem{pbh4} M. E. Rodrigues and Z. A. A. Oporto, Phys. Rev. D {\bf85} 104022 (2012), arXiv:1201.5337.

\bibitem{pbh5} D. F. Jardim, M. E. Rodrigues, and  M. J. S. Houndjo, Eur. Phys. J. Plus {\bf127}, 123 (2012), arXiv:1202.2830.

\bibitem{pbh6} A. Nakonieczna, M. Rogatko and R. Moderski, Phys. Rev. D {\bf86}, 044043  (2012), arXiv:1209.1203.



\bibitem{pbh8} S. V. Bolokhov, K. A. Bronnikov and M. V. Skvortsova,  Class. Quant. Grav. {\bf29}, 245006 (2012).

\bibitem{pbh9} S. Chen and J. Jing, Class. Quantum Grav. {\bf30}, 175012 (2013).

\bibitem{pbh7} M. Azreg-Ainou,  Phys. Rev. D {\bf87}, 024012 (2013), arXiv:1209.5232.

\bibitem{GL1} C. Ding, C. Liu, Y. Xiao, L. Jiang and R. G. Cai, Phys. Rev. D {\bf88}, 104007 (2013).
\bibitem{GL2} E. F. Eiroa and C. M. Sendra, Phys. Rev. D {\bf88}, 103007 (2013).
\bibitem{GL3}  G. N. Gyulchev and I. Z. Stefanov, Phys. Rev. D {\bf87},063005 (2013).
\bibitem{GL4} M. Sharif and S. Iftikhar,   Adv. High Energy Phys. {\bf 2015}, 854264 (2015).

\bibitem{qpo1} R. A. Remillard and J. E. McClintock, Ann. Rev. Astron.
Astrophys. {\bf44}, 49 (2006).
\bibitem{qpo2} T. M. Belloni and S. E. Motta, arXiv:1603.07872.


\bibitem{RPM1} S. E. Motta, T. M. Belloni, L. Stella, T. Muoz-Darias
and R. Fender, Mon. Not. Roy. Astron. Soc. {\bf437}, 2554
(2014) [arXiv:1309.3652 [astro-ph.HE]].

\bibitem{RPM2} S. E. Motta, T. Muoz-Darias, A. Sanna, R. Fender,
T. Belloni and L. Stella, Mon. Not. Roy. Astron. Soc.
{\bf439}, 65 (2014) [arXiv:1312.3114 [astro-ph.HE]].
\bibitem{RPM20} P. Casella,  T. Belloni, and L. Stella, L. Astrophys. J. {\bf629}, 403 (2005).

\bibitem{RPM3} L. Stella and M. Vietri, Astrophys. J. {\bf492}, L59 (1998)
[astro-ph/9709085].
\bibitem{RPM31} L. Stella and M. Vietri, Phys. Rev.
Lett. {\bf82}, 17 (1999) [astro-ph/9812124].
\bibitem{RPM32} L. Stella, M. Vietri and S. Morsink, Astrophys. J. {\bf524}, L63 (1999) [astro-ph/9907346].

\bibitem{RMP4} A. Ingram, C. Done, and P. C. Fragile, Mon. Not. Roy. Astron. Soc. {\bf397}, L101 (2009).
\bibitem{RMP5} P. C. Fragile, O. Straub, and O. Blaes, 	arXiv:1602.08082.

\bibitem{TB1} C. Bambi, Eur. Phys. J. C {\bf75}, 162 (2015).
\bibitem{TB11} C. Bambi and S. Nampalliwar, arXiv:1604.02643.
\bibitem{TB2} Z. Stuchlik and A. Kotrlova, Gen. Rel. Grav. {\bf41}, 1305
(2009).
\bibitem{TB3} T. Johannsen and D. Psaltis, Astrophys. J. {\bf726}, 11
(2011) [arXiv:1010.1000 [astro-ph.HE]].

\bibitem{TB5} A. Maselli, L. Gualtieri, P. Pani, L. Stella, and V. Ferrari, Astrophys. J. {\bf801}, 115 (2015).

\bibitem{TB6}A. G. Suvorov and A. Melatos,  Phys. Rev. D {\bf93}, 024004 (2016).
\bibitem{TB7}G. Pappas, Mon. Not. R. Astron. Soc. {\bf422}, 2581-2589 (2012).

\bibitem{TB8} K. Boshkayev, D. Bini, J. Rueda, A. Geralico, M. Muccino and I. Siutsou, Grav. Cosmol. {\bf20}, 233-239 (2014).

\bibitem{NJ1}E. T. Newman and A. I. Janis, J. Math. Phys. (N.Y.) {\bf6}, 915
(1965).

\bibitem{TB4} M. E. Beer and P. Podsiadlowski, Mon. Not. Roy. Astron.
Soc. {\bf331}, 351 (2002) [astro-ph/0109136].
\end{thebibliography}
\end{document}